\input harvmac
%\draftmode
\let\includefigures=\iftrue
\let\useblackboard=\iftrue
\newfam\black

%Figure Stuff
\includefigures
\message{If you do not have epsf.tex (to include figures),}
\message{change the option at the top of the tex file.}
\input epsf
\def\figin{\epsfcheck\figin}\def\figins{\epsfcheck\figins}
\def\epsfcheck{\ifx\epsfbox\UnDeFiNeD
\message{(NO epsf.tex, FIGURES WILL BE IGNORED)}
\gdef\figin##1{\vskip2in}\gdef\figins##1{\hskip.5in}% blank space instead
\else\message{(FIGURES WILL BE INCLUDED)}%
\gdef\figin##1{##1}\gdef\figins##1{##1}\fi}
\def\DefWarn#1{}
\def\figinsert{\goodbreak\midinsert}
\def\ifig#1#2#3{\DefWarn#1\xdef#1{fig.~\the\figno}
\writedef{#1\leftbracket fig.\noexpand~\the\figno}%
\figinsert\figin{\centerline{#3}}\medskip\centerline{\vbox{
\baselineskip12pt\advance\hsize by -1truein
\noindent\footnotefont{\bf Fig.~\the\figno:} #2}}
%\bigskip
\endinsert\global\advance\figno by1}
%%%
\else
\def\ifig#1#2#3{\xdef#1{fig.~\the\figno}
\writedef{#1\leftbracket fig.\noexpand~\the\figno}%
%\figinsert\figin{\centerline{#3}}\medskip
%\centerline{\vbox{\baselineskip12pt
%\advance\hsize by -1truein\noindent
%\footnotefont{\bf Fig.~\the\figno:} #2}}
%\bigskip\endinsert
\global\advance\figno by1} \fi

\def\id{{1 \kern-.28em {\rm l}}}

\def\K3{{\bf K3}}
\def\journal#1&#2(#3){\unskip, \sl #1\ \bf #2 \rm(19#3) }
\def\andjournal#1&#2(#3){\sl #1~\bf #2 \rm (19#3) }

\def\bar{\overline}

\def\frac#1#2{{#1\over#2}}

\def\inbar{\,\vrule height1.5ex width.4pt depth0pt}
\def\IC{\relax\hbox{$\inbar\kern-.3em{\rm C}$}}
\def\IR{\relax{\rm I\kern-.18em R}}
\def\IP{\relax{\rm I\kern-.18em P}}

%
%%%%%%%%%%%%%%%%%%%%%%%%%%%%%%%%%%%%
%

%
\catcode`\@=11
\def\slash#1{\mathord{\mathpalette\c@ncel{#1}}}
\overfullrule=0pt

\def\OO{{\cal O}}

\def\underrel#1\over#2{\mathrel{\mathop{\kern\z@#1}\limits_{#2}}}

\catcode`\@=12

%%%%%%%%%%%%%%%%%%%%%%%%%%%%%%%%%%%%%%%%%%%%%%%%%%%%%%%%%%%%%%

%

%%%%%%%%%%%%%%%%%%%%%%%%%%%%%%%%%%%%%%%%%%%%%%%%%%%%%%%%%%%%%%
% new defs:

\def\p{{\partial}}

\def\ra{{\rightarrow}}

%\SakaiCN
\lref\SakaiCN{
T.~Sakai and S.~Sugimoto,
``Low energy hadron physics in holographic QCD,''
Prog.\ Theor.\ Phys.\  {\bf 113}, 843 (2005)
[arXiv:hep-th/0412141].
%%CITATION = HEP-TH 0412141;%%
%%Cited 33 times in SPIRES-HEP
}

%\SonET
\lref\SonET{
  D.~T.~Son and M.~A.~Stephanov,
  ``QCD and dimensional deconstruction,''
  Phys.\ Rev.\ D {\bf 69}, 065020 (2004)
  [arXiv:hep-ph/0304182].
  %%CITATION = HEP-PH 0304182;%%
}

%\BabingtonVM
\lref\BabingtonVM{
  J.~Babington, J.~Erdmenger, N.~J.~Evans, Z.~Guralnik and I.~Kirsch,
  ``Chiral symmetry breaking and pions in non-supersymmetric gauge /  gravity
  duals,''
  Phys.\ Rev.\ D {\bf 69}, 066007 (2004)
  [arXiv:hep-th/0306018].
  %%CITATION = HEP-TH 0306018;%%
}
%\EvansIA
\lref\EvansIA{
  N.~J.~Evans and J.~P.~Shock,
  ``Chiral dynamics from AdS space,''
  Phys.\ Rev.\ D {\bf 70}, 046002 (2004)
  [arXiv:hep-th/0403279].
  %%CITATION = HEP-TH 0403279;%%
}
%\GhorokuSP
\lref\GhorokuSP{
  K.~Ghoroku and M.~Yahiro,
  ``Chiral symmetry breaking driven by dilaton,''
  Phys.\ Lett.\ B {\bf 604}, 235 (2004)
  [arXiv:hep-th/0408040].
  %%CITATION = HEP-TH 0408040;%%
}

%\ErlichQH
\lref\ErlichQH{
  J.~Erlich, E.~Katz, D.~T.~Son and M.~A.~Stephanov,
  ``QCD and a holographic model of hadrons,''
  Phys.\ Rev.\ Lett.\  {\bf 95}, 261602 (2005)
  [arXiv:hep-ph/0501128].
  %%CITATION = HEP-PH 0501128;%%
}

%\KarchPV
\lref\KarchPV{
  A.~Karch, E.~Katz, D.~T.~Son and M.~A.~Stephanov,
  ``Linear confinement and AdS/QCD,''
  arXiv:hep-ph/0602229.
  %%CITATION = HEP-PH 0602229;%%
}

\lref\AHJK{
  E.~Antonyan, J.~A.~Harvey, S.~Jensen and D.~Kutasov,
  ``NJL and QCD from string theory,''
  arXiv:hep-th/0604017.
  %%CITATION = HEP-TH 0604017;%%
}

%\WittenZW
\lref\WittenZW{
  E.~Witten,
  ``Anti-de Sitter space, thermal phase transition, and confinement in  gauge
  theories,''
  Adv.\ Theor.\ Math.\ Phys.\  {\bf 2}, 505 (1998)
  [arXiv:hep-th/9803131].
  %%CITATION = HEP-TH 9803131;%%
}

%\HawkingDH
\lref\HawkingDH{
  S.~W.~Hawking and D.~N.~Page,
  ``Thermodynamics Of Black Holes In Anti-De Sitter Space,''
  Commun.\ Math.\ Phys.\  {\bf 87}, 577 (1983).
  %%CITATION = CMPHA,87,577;%%
}

%\NambuTP
\lref\NambuTP{Y.~Nambu and G.~Jona-Lasinio,
``Dynamical Model Of Elementary Particles Based On An Analogy With
Superconductivity. I,''Phys.\ Rev.\  {\bf 122}, 345 (1961).
%%CITATION = PHRVA,122,345;%%
}

%\HatsudaPI
\lref\HatsudaPI{
T.~Hatsuda and T.~Kunihiro,
``QCD phenomenology based on a chiral effective Lagrangian,''
Phys.\ Rept.\  {\bf 247}, 221 (1994)
[arXiv:hep-ph/9401310].
%%CITATION = HEP-PH 9401310;%%
}

%\VolkovKW
\lref\VolkovKW{
M.~K.~Volkov and A.~E.~Radzhabov,
``Forty-fifth anniversary of the Nambu-Jona-Lasinio model,''
arXiv:hep-ph/0508263.
%%CITATION = HEP-PH 0508263;%%
}

\lref\KlevanskyQE{
  S.~P.~Klevansky,
  ``The Nambu-Jona-Lasinio model of quantum chromodynamics,''
  Rev.\ Mod.\ Phys.\  {\bf 64}, 649 (1992).
  %%CITATION = RMPHA,64,649;%%
}

\lref\BuballaQV{
  M.~Buballa,
  ``NJL model analysis of quark matter at large density,''
  Phys.\ Rept.\  {\bf 407}, 205 (2005)
  [arXiv:hep-ph/0402234].
  %%CITATION = HEP-PH 0402234;%%
}
%\GrossJV
\lref\GrossJV{
D.~J.~Gross and A.~Neveu,
``Dynamical Symmetry Breaking In Asymptotically Free Field Theories,''
Phys.\ Rev.\ D {\bf 10}, 3235 (1974).
%%CITATION = PHRVA,D10,3235;%%
}

%\GiveonSR
\lref\GiveonSR{
A.~Giveon and D.~Kutasov,
``Brane dynamics and gauge theory,''
Rev.\ Mod.\ Phys.\  {\bf 71}, 983 (1999)
[arXiv:hep-th/9802067].
%%CITATION = HEP-TH 9802067;%%
}

%\SakaiYT
\lref\SakaiYT{
  T.~Sakai and S.~Sugimoto,
  ``More on a holographic dual of QCD,''
  Prog.\ Theor.\ Phys.\  {\bf 114}, 1083 (2006)
  [arXiv:hep-th/0507073].
  %%CITATION = HEP-TH 0507073;%%
}

%\ItzhakiDD
\lref\ItzhakiDD{
N.~Itzhaki, J.~M.~Maldacena, J.~Sonnenschein and S.~Yankielowicz,
``Supergravity and the large N limit of theories with sixteen
supercharges,''
Phys.\ Rev.\ D {\bf 58}, 046004 (1998)
[arXiv:hep-th/9802042].
%%CITATION = HEP-TH 9802042;%%
}

%\KruczenskiUQ
\lref\KruczenskiUQ{
  M.~Kruczenski, D.~Mateos, R.~C.~Myers and D.~J.~Winters,
  ``Towards a holographic dual of large-N(c) QCD,''
  JHEP {\bf 0405}, 041 (2004)
  [arXiv:hep-th/0311270].
  %%CITATION = HEP-TH 0311270;%%
}

%\KarchSH
\lref\KarchSH{
  A.~Karch and E.~Katz,
  ``Adding flavor to AdS/CFT,''
  JHEP {\bf 0206}, 043 (2002)
  [arXiv:hep-th/0205236].
  %%CITATION = HEP-TH 0205236;%%
}

%\StephanovWX
\lref\StephanovWX{
  M.~A.~Stephanov,
  ``QCD phase diagram and the critical point,''
  Prog.\ Theor.\ Phys.\ Suppl.\  {\bf 153}, 139 (2004)
  [Int.\ J.\ Mod.\ Phys.\ A {\bf 20}, 4387 (2005)]
  [arXiv:hep-ph/0402115].
  %%CITATION = HEP-PH 0402115;%%
}

\lref\ASY{
  O.~Aharony, J.~Sonnenschein and S.~Yankielowicz,
  ``A holographic model of deconfinement and chiral symmetry restoration,''
  arXiv:hep-th/0604161.
  %%CITATION = HEP-TH 0604161;%%
}

\Title{\vbox{\baselineskip12pt
\hbox{hep-th/0604173}
}}
{\vbox{\centerline{Chiral Phase Transition from String Theory}
\vskip.06in
}}
\centerline{Andrei Parnachev and David A. Sahakyan}
\bigskip
\centerline{{\it Department of Physics, Rutgers University}}
\centerline{\it Piscataway, NJ 08854-8019, USA}
 \vskip.1in \vskip.1in \centerline{\bf Abstract}  
\noindent

The low energy dynamics of a certain D-brane configuration in string theory 
is described at weak t'Hooft coupling
by a non-local version of the Nambu-Jona-Lasinio model.
We study this system at finite temperature and strong t'Hooft coupling,
using the string theory dual.
We show that for sufficiently low temperatures chiral symmetry 
is broken, while for temperatures larger then the critical value, it gets restored.
We compute the latent heat and observe that  
the phase transition is of the first order.

\vfill

\Date{April 24, 2006}
   
%\draftmode

\newsec{Introduction}

Over the past few years there has been considerable progress in studying
QCD-like theories using holography (an incomplete list of references includes
\refs{\KarchSH\SonET\BabingtonVM\KruczenskiUQ\SakaiCN
\ErlichQH-\SakaiYT}).
The gravity description is useful for studying  properties of the hadrons
as well as  qualitative features of 
QCD such as confinement and chiral symmetry breaking ($\chi$SB).
%One of the important features of QCD observed at low energies 
%is confinement.
%Another is chiral symmetry breaking
The latter  stands for the dynamical
breaking of the $U(N_f)\times U(N_f)$ symmetry acting on $N_f$ right 
handed and left handed quarks of the high energy  Lagrangian down
to its diagonal subgroup, $U(N_f)$.
%To complicate things, 
In QCD the energy scales of confinement and $\chi$SB
are approximately the same.

An interesting example which 
exhibits chiral symmetry breaking is the Nambu-Jona-Lasinio (NJL) model \NambuTP,
%(for recent reviews see \eg\ \refs{\HatsudaPI,\BuballaQV})
whose Lagrangian contains left and right-handed quarks interacting via 
a non-renormalizable quartic term.
The non-renormalizability of this interaction makes it natural to attempt a 
well-defined UV completion.
Recently, such a model was proposed within the context of string theory \AHJK.
The weak coupling string theory description of the model 
considered in \AHJK\ involves a set of $N_f$ D8-branes and $N_f$ anti-D8-branes
separated by a distance $L$ in the $x^4$ direction and $N_c$ D4-branes
extended in the $(x^0,\ldots,x^4)$ directions.
At low energies, the system reduces to the $N_f$ left handed quarks 
and $N_f$ right handed quarks interacting via a five-dimensional gluon exchange.
This induces a non-local Nambu-Jona-Lasinio-type interaction in the low
energy Lagrangian for the quarks.
%It has been shown that chiral symmetry is dynamically broken both at weak and strong
%t'Hooft coupling.

An important feature of the large $N_c$  model studied in \AHJK\ is the separation
of scales. Namely, the energy scale associated with confinement is well below
that of chiral symmetry breaking.
In addition, by compactifying the $x^4$ direction and varying the radius 
of the resulting circle one can smoothly interpolate between the NJL
model and QCD.
The weakly coupled region of the former can be analyzed within field theory
%The effective potential for the quark condensate yields the gap equation, which 
%features momentum-dependent solution that  breaks chiral symmetry for 
%arbitrarily weak coupling 
\AHJK.
At strong coupling, one can use holographically dual string theory.
Chiral symmetry breaking and meson properties can be analyzed within
the DBI action for the $N_f$ D8 branes living in the background created by $N_c$
D4 branes.
As long as $N_f<<N_c$, back reaction of the D8-brane on the background metric
can be ignored.
%(Recent work along these lines include \refs{\KarchSH\KruczenskiUQ\SakaiCN-\SakaiYT}.)
The conclusion of \AHJK\ is that at strong t'Hooft coupling chiral symmetry
is dynamically broken, since the corresponding brane configuration is energetically
favored over the one with unbroken symmetry.

In this paper we study the effects of finite temperature in the limit
of strong t'Hooft coupling.
We find that there is a phase transition at the temperature $T_c\approx 0.15 L^{-1}$.
Below this temperature, there exists a thermodynamically
favorable brane configuration where chiral symmetry is dynamically broken.
As we increase the temperature above the $T_c$, we find that the solution with the
unbroken chiral symmetry is thermodynamically preferred. 
As temperature reaches $T_*\approx 0.17 L^{-1}$
the phase with broken chiral symmetry ceases to exist. 
These properties suggest that the phase transition is of the first order.
To confirm this we compute the latent heat, and show that it is positive.
%When the temperature exceeds $T_c$ this solution is no longer
%available, and chiral symmetry gets restored.
The rest of the paper is organized as follows.
In the next section we introduce the brane model at finite temperature
and discuss the features of the phase transition at $T=T_c$.
We discuss our results in Section 3.

Note added: As we were completing this paper we learned of the
paper by O. Aharony, C. Sonnenschein and S. Yankielowicz \ASY\ which
partially overlaps with our results.

%%%%%%%%%%%%%%%%%%%%%%%%%%%%%%%%%%%%%%%%%%%%%%%%%%%

\newsec{D8/D4 branes at finite temperature}

We will consider the near-horizon geometry of the D4-branes in the
limit studied in \AHJK, where the direction transverse to the D8 
branes is non-compact.
The near-horizon geometry of the $D4$-branes at finite temperature is: 
\eqn\metricss{ ds^2=\left(U\over R\right)^{3\over2}\left(dx_i dx^i+f(U)dt^2+(dx^4)^2\right)+
\left(U\over R\right)^{-{3\over2}}\left({dU^2\over f(U)}+U^2d\Omega_4^2\right)} 
where $f(U)=1-U_{T}^3/U^3$.
Here $t$ is the Euclidean time, $i=1,\ldots,3$; $U$ and $\Omega_4$ label the
radial and angular directions of $(x^5,\ldots,x^9)$. The parameter $R$ is given by
\eqn\rdefn{R^3 = \pi g_s N_c l_s^3=\pi\lambda}
where in the last equality and in the rest of the paper we set $\alpha'=1$.
The fourbrane geometry also has a non-trivial dilaton background,
\eqn\dfourdil{e^\Phi=g_s\left(U\over R\right)^{3\over4}}
Finite temperature implies that $t$ is a periodic variable, 
\eqn\tper{ t\sim t+\beta  }
On the other hand, in order for \metricss\ to describe a non-singular space,
$t$ must satisfy
\eqn\xfourpd{ t=t+{4\pi R^{3/2}\over 3 U_{T}^{1/2}}   }
Hence, the temperature is related to the minimal value of $U$,
denoted by $U_T$, as
$T=3 U_{T}^{1/2}/4\pi R^{3/2}$.

Quarks can be incorporated into this system by adding
$N_f$ $D8-\bar D8$ brane pairs.
In the following we consider $N_f=1$ case for simplicity, 
but the discussion applies for finite $N_f<<N_c$ as well.
At weak coupling, $D8$ and $\bar D8$ branes are separated by
the distance $L$ in the $x^4$ direction.
At strong coupling $\lambda>>L,l_s$, 
the shape of $D8-\bar D8$ brane pair is determined by the equation
of motion that follows from the DBI action.
As we will see below, there are two possible types of solutions, straight
and curved.
The straight solution is simply $\tau=const$, and describes separated
$D8$ and $\bar D8$ branes.
In this situation, chiral symmetry is not broken.
In the Lorentzian version of \metricss, the point $U=U_T$ corresponds
to the horizon of the black hole.
The straight branes therefore cross the horizon in the Lorentzian space.
The second type of solution is the one where the $D8-\bar D8$ brane surface
is connected and is described by a curve $U(x_4)$ in the 
geometry \metricss.
This configuration breaks chiral symmetry and does not cross the horizon
in the Lorentzian space.
To understand which configuration is thermodynamically preferred, it is necessary
to compare the values of the free energy, which we do below.

In the following it will be convenient to parameterize the $D8-\bar D8$
brane surface by the
embedding function $\tau(x^\mu,U)\equiv x^4(x^\mu,U)$.
The classical trajectory is $x^\mu$ independent, and
$U\ra\infty$ as $\tau\ra\pm L/2$ \AHJK.
The D-brane has two branches which are interchanged by $\tau\ra-\tau$.
We will mostly restrict attention to one branch,
where $U$ varies from $\infty$ to the limiting value $U_0$ at $\tau=0$.
The induced metric on the brane in the background \metricss\ is
\eqn\indmet{
\eqalign{ ds^2&=\left({U\over R}\right)^{3/2}
  \left((\eta_{\mu\nu}+\p_\mu\tau\p_\nu\tau)dx_\mu dx^\nu+(f(U)-1)dt^2\right)
    +2 \left(U\over R\right)^{3/2} \p_U\tau \p_\mu \tau dx^\mu d U \cr
    &\qquad+\left({U\over R}\right)^{-3/2}\left({1\over f(U)}+\left(U\over R\right)^{3} 
   (\p_U\tau)^2\right) dU^2
    +R^{3/2} U^{1/2} d\Omega_4^2   \cr}
}
where $\mu=0,\ldots,3$.
At this point we assume that the brane configuration is 
independent of $x^\mu$.
The DBI action
%, up to an overall numerical coefficient, 
is
\eqn\dbitu{  S={T_8 V_S R^{3/2}\over T}\int d^3 x
               \int dU U^{5/2}\sqrt{1+(U/R)^3 f(U)(\p_U\tau)^2}   } 
%The equation of motion that follows from the DBI action is
%\eqn\eomtu{  \left({d\tau\over dU}\right)^2={R^3\over U^{11} (U_0^{-8}-U^{-8})  } }
%
where $T_8=1/(2\pi)^8g_s$ is the tension of the brane and $V_S=\pi^2/2$ is the volume of the unit four-sphere.
It is convenient to make a change of variables
\eqn\uy{  y={2 R^{3/2}\over\sqrt{U}},\qquad U={4 R^3\over y^2}  }
%\eqn\utoy{  \left({R\over U}\right)^{3/2} dU=dy    }
which brings \dbitu\ to the form 
\eqn\dbity{  S={2^{6} \pi^2 T_8 R^{12}\over T}\int d^3 x
             \int dy y^{-8} \sqrt{ 1+ f(y) (\p_y\tau)^2}   }
where 
\eqn\fofy{  f(y)=1-{y^6\over y_T^6},\qquad y_T={3\over2\pi T}       } 
The space is restricted to lie in the region $y\in(0,y_T)$, whose
upper limit corresponds to the black hole horizon which shrinks to a point
after the Wick rotation.
The equation of motion in the new variables is
\eqn\eoma{  \p_y\left[ {y^{-8} f(y) \p_y\tau\over\sqrt{1+f (\p_y\tau)^2}}\right]=0  }
We denote conserved quantity in the square brackets by $y_0^{-8}$;
this parameter parameterizes the solution.
Then, \eoma\ can be written in the form
\eqn\eomty{  \left({d\tau\over dy}\right)^2={1\over f(y)(f(y) y_0^{16}/y^{16}-1)} }
Note that the brane extends all the way from $y=0$ (which corresponds 
to the asymptotic region, $U\ra\infty$) to $y=y_*$ which is determined
by
\eqn\ystar{   (1-{y_*^6\over y_T^6})  {y_0^{16}\over y_*^{16}}-1=0   }
In particular, for $y_T>>y_0$, $y_*\approx y_0$, while in the
opposite regime $y_T<<y_0$, 
\eqn\ystarlim{  y_*=y_T\left(1-{y_T^{16}\over6 y_0^{16}}+\OO(y_T/y_0)^{32}\right)   }
In the limit $y_0\ra\infty$ the D-brane solution gets closer and closer
to the horizon.

In the discussion above $y_0$ is a constant which parameterizes the solutions
of equation of motion.
Its relation to the physical parameters $y_T=3/2\pi T$
and $L$ is determined by
\eqn\intl{   {L\over2}=\int_0^{y_*} {dy\over \sqrt{ f(y)(f(y) y_0^{16}/y^{16}-1) } }=
                  y_T \int_0^{x_*} {dx\over \sqrt{ f(x)(f(x) x_0^{16}/x^{16}-1) } }     } 
where we introduced the rescaled variable $x=y/y_T$.
Analogously, $x_0=y_0/y_T$ and $x_*=y_*/y_T$.
In the limit of small $y_0$, $L$  differs from it by a multiplicative factor
of order unity,  
\eqn\lyolt{  L\approx y_0 { B(9/16,1/2)\over8},\qquad y_0<<y_T }
In the opposite regime we need to estimate the integral,
\eqn\intest{   {L\over2}\approx  y_T \int_0^{x_*} 
      {dx\over\sqrt{(1-x^6)((1-x^6)x^{16}/x_0^{16}-1)}},\qquad
x_0\ra\infty   }
To estimate the behavior of the integral in \intest\ at large $x_0$
it is convenient to introduce a new variable
\eqn\newvarz{  z=(1-x^6)-{x^{16}\over x_0^{16}}   }
omitting terms subleading in $1/x_0^{16}$ we obtain
\eqn\intestz{  L\approx {y_T\over 3} \int_0^1 {dz\over x^5\sqrt{z}\sqrt{x^{16}+x_0^{16} z}  }
    \sim y_T \left({y_T\over y_0}\right)^8 }
which goes rapidly to zero as $y_0\ra\infty$.
\midinsert\bigskip{\vbox{{\epsfxsize=3in
        \nobreak
    \centerline{\epsfbox{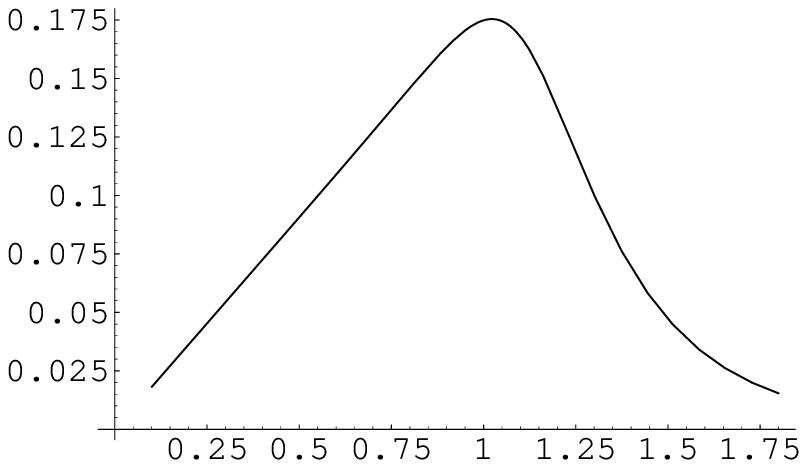}}
        \nobreak\bigskip
    {\raggedright\it \vbox{
{\bf Fig 1.}
{\it  The value of $L/2$ in the units of $y_T$ [eq. \intl] as a function of $x_0=y_0/y_T$.
}}}}}}
\bigskip\endinsert
We are thus led to the following picture.
The integral over $x$ in \intl\ goes like $x_0$ for small $x_0$ and
peaks around $x_0\sim 1$.
It then goes down and goes rapidly to zero as $x_0\ra\infty$.
Hence, for a sufficiently small temperature $T<<1/L$ we generically have
two solutions for a D8 brane which asymptotes to $x^4=\pm L/2$.
This picture is confirmed by the numerical evaluation of the
integral in \intl\ shown in Fig. 1.

For a sufficiently high temperature, there are no curved solutions,
although there still exists a solution with constant $x^4$.
In Minkowski space, this solution describes two collections of branes and
anti-branes which cross the horizon.
For this configuration chiral symmetry is restored, which is what we expect
from the high temperature phase.
To understand the details of the chiral symmetry restoration, we need
to compare the values of the action for the solutions described above.

The difference between the values of the free energy for the straight 
and curved solutions is
\eqn\diffac{\eqalign{
&{1\over T}({\cal F}_{straight}-{\cal F}_{curved})=\delta S=S_{straight}-S_{curved}\propto\cr
&\int_{y_*}^{y_T} dy (y^{-8}-0)
   +\int_0^{y_*} dy y^{-8} 
  \left( 1-\left(1-{y^{16}\over y_0^{16}(1-y^6/y_T^6)}\right)^{-1/2}
    \right)     }}
When the temperature is small, $y_T>>L$, it is easy to consider
the branch with $y_0\sim L$.
There, the result is the same
as in \AHJK, 
\eqn\dsa{ \delta S\sim y_0^{-7}\sim y_T^{-7} (y_T/L)^7   }
In particular, it is positive, i.e. the curved solution is preferred and
chiral symmetry is unbroken.
Moreover, $\delta S$ blows up  as $L\ra0$.
\midinsert\bigskip{\vbox{{\epsfxsize=3in
        \nobreak
    \centerline{\epsfbox{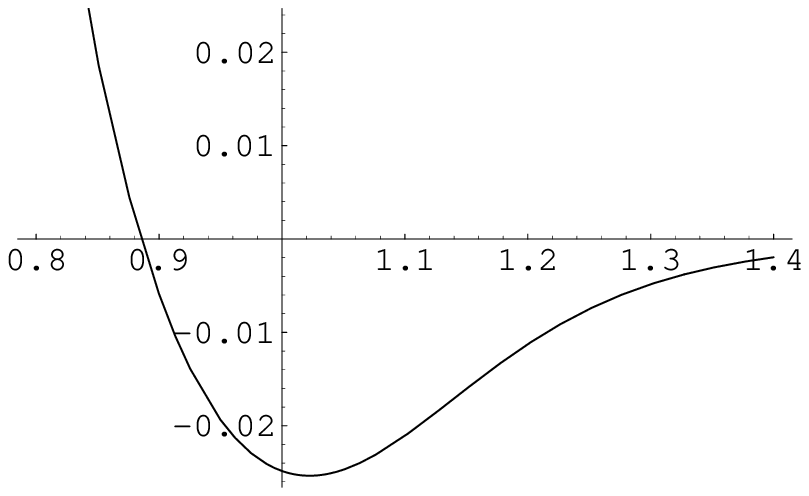}}
        \nobreak\bigskip
    {\raggedright\it \vbox{
{\bf Fig 2.}
{\it  The value of $\delta S$ in appropriate units [eq. \diffac] as a function of $x_0=y_0/y_T$.
}}}}}}
\bigskip\endinsert
Consider now the second branch.
Now the integral in \diffac\ is dominated near the region $y\approx y_*$.
To see this, introduce a new variable $z$ defined by \newvarz.
The relevant part of the integral in \diffac\ is
\eqn\reldc{  \int_0^1 (1-\sqrt{1+{1\over x_0^{16} z}})\sim- x_0^{-16}\log x_0 }
where we neglect terms that are $\OO(x_0^{-16})$
The result therefore is negative, and goes to zero as $L\ra 0$:
\eqn\dsb{  \delta S\sim -y_T^{-7} (y_T/y_0)^{-16} \log(y_0/y_T)\sim -y_T^{-7} (L/y_T)^2\log L }
where we used \intestz.
In fact, as we see on Fig. 2, the difference \diffac\ 
is positive for sufficiently small $y_0$, but then becomes negative.

The picture therefore  is the following.
When the temperature is low compared to $1/L$, the system resides in the
state where chiral symmetry is broken and $y_0\sim L$.
This state is described by the graph in Fig. 1 with $x_0<<1$.
From Fig. 2 it is clear that the action for this branch is smaller 
then that for the straight brane solution, where chiral symmetry 
is restored.
It is interesting that there exists a second solution, with large $y_0$
for a sufficiently low temperature.
This solution also breaks chiral symmetry, but is not thermodynamically
preferred.

When the temperature is raised above the critical temperature
$T_c\approx 0.15 L^{-1}$, the system undergoes a phase transition:
the straight brane becomes thermodynamically preferred, and chiral
symmetry gets restored.
(The critical temperature corresponds to $x_0\approx 0.885$, according to
Fig. 2. This translates into $L/2 y_T\approx 0.161$, according to Fig. 1.
Using \fofy\ gives the value of $T_c$ quoted above).
As the temperature is increased past $T_*\approx 0.17 L^{-1}$,
the only available solution is a straight brane.
This picture suggests that the transition is of first order.
Indeed it is easy to see that the latent heat is non-zero at $T=T_c$.
The simplest way to see this is to note that  the entropy density jumps at the transition
\eqn\entr{{\Delta{\cal S}\over V_3}=-{1\over V_3}{\p({\cal F}_{straight}-{\cal F}_{curved})\over\p T }}
Using \diffac, and computing the derivative at $T=T_c$ we find
\eqn\entrone{
{\Delta{\cal S}\over V_3}\simeq 2.16 T_8 V_S \lambda^4 \left({4\pi\over 3}\right)^7 T_c^6
%x_*(T_c) {\p I[x_*(T_c)]\over\p x_*}
}
The latent heat per unit volume is positive and equal to
\eqn\latheat{C_l={T_c({\cal S}_{straight}-{\cal S}_{curved})\over V_3}\simeq 0.03 \pi^5 N_c \lambda^3 T_c^7}

\newsec{Discussion}

In this paper we considered the string theory description of the chiral symmetry 
phase transition.
The corresponding D-brane system at weak coupling is described by a certain NJL model
with non-local interaction of the quarks.
We argued that when the system is strongly coupled, $\lambda>>L$,
the following scenario takes  place.
As the temperature is low compared to the scale set by the brane separation,
$1/L$, the curved D8-brane solution which breaks chiral symmetry
is thermodynamically preferred over the straight D8-brane solution, 
with unbroken chiral symmetry.
This solution goes over smoothly to the zero temperature solution
studied in \AHJK.
In particular, the maximal value of $y$ on the D8-brane trajectory,
denoted $y_*$ in the paper, scales as $y_*\sim L$ for small $L$,
i.e. the D-brane is getting further away from the horizon at $y=y_T$
as $L$ decreases.
As the temperature approaches $T_c\approx 0.15 L^{-1}$ from below, the free energy
of the straight brane solution approaches that of the curved one.
At $T=T_c$ the derivative of the free energy with respect to the temperature jumps,
which indicates the first order phase transition.
When the temperature is raised above $T_c\approx 0.15 L^{-1}$, the
straight brane solution, which restores chiral symmetry becomes
thermodynamically preferred.
For temperatures higher then  $T_*\approx 0.17 L^{-1}$,
the straight brane is the only available solution. 
At low temperatures 
%(equivalently, at low $L<<1/T$) 
there also exists a second curved D8-brane solution,
which comes asymptotically close to the horizon as $L\ra0$.
The value of the action for this solution is higher
then that of the other two.

To reiterate, the scale of the temperature where chiral phase transition
happens is set by the asymptotic separation of the D8-branes $L$.
It is interesting that this energy scale is much larger then 
that of Hawking-Page transition which is associated with confinement/deconfinement
transition in field theory.
To understand this better, consider the space \metricss\ where
the $x^4$ direction is compactified on a circle of radius $R_4$ \SakaiCN.
In this case, the scale of Hawking-Page transition is set by $y_T\sim R_4$.
The case of $R_4\sim L$ corresponds to QCD, where
both confinement and chiral symmetry breaking happen at the same scale.
On the other hand in the case $R_4>>L$, where
our
analysis applies, the two scales are widely separated, as has been also
argued
in \AHJK.

There are many open questions.
It would be interesting to generalize the discussion of this 
paper for a non-zero chemical potential. 
There one expects non-trivial phase structure in the $(\mu,T)$ plane.
%(see \eg\ \StephanovWX).
Understanding the behavior of the mesons at finite temperature
is another direction.
%Finally, it would be interesting to understand the phase transition
%at weak coupling, generalizing 
%field theory results of \AHJK\ to finite temperature. 

\bigskip
{\noindent\bf Acknowledgements:}
We would like to thank E. Antonyan, M. Douglas, J. Harvey, D. Kutasov and A. Ryzhov 
for discussions and correspondence.
A.P. is grateful to the Theory Group at the University of Chicago for hospitality.
This work was supported in part by  DOE grant DE-FIG02-96ER40949.

\listrefs
\end